# Origin of current-controlled negative differential resistance modes and the emergence of composite characteristics with high complexity

Shuai Li[1], Xinjun Liu[2*], Sanjoy Kumar Nandi[1], Shimul Kanti Nath[1], Robert G. Elliman[1*]

[1]Department of Electronic Materials Engineering, Research School of Physics and Engineering, The Australian National University, Canberra ACT 2601, Australia

[2]Tianjin Key Laboratory of Low Dimensional Materials Physics and Preparation Technology, Faculty of Science, Tianjin University, Tianjin 300354, China

*Address correspondence to xinjun.liu@tju.edu.cn and rob.elliman@anu.edu.au

## Abstract

Current-controlled negative differential resistance has significant potential as a fundamental building block in brain-inspired neuromorphic computing. However, achieving desired negative differential resistance characteristics, which is crucial for practical implementation, remains challenging due to little consensus on the underlying mechanism and unclear design criteria. Here, we report a material-independent model of current-controlled negative differential resistance to explain a broad range of characteristics, including the origin of the discontinuous snap-back response observed in many transition metal oxides. This is achieved by explicitly accounting for a non-uniform current distribution in the oxide film and its impact on the effective circuit of the device, rather than a material-specific phase transition. The predictions of the model are then compared with experimental observations to show that the continuous S-type and discontinuous snap-back characteristics serve as fundamental building blocks for composite behaviour with higher complexity. Finally, we demonstrate the potential of our approach for predicting and engineering unconventional compound behaviour with novel functionality for emerging electronic and neuromorphic computing applications.

## Introduction

Current-controlled negative differential resistance (NDR) is of increasing interest for brain-inspired computing based on a number of promising applications, including trigger comparators[1], self-sustained and chaotic oscillators[2-7], threshold logic devices[8,9] and the emulation of biological neuronal dynamics[10,11]. Since the functionality of such devices is determined by the specific form of their current-voltage characteristics, it is important to understand the physical basis of switching and how it is affected by device structure and operating conditions. This is complicated by the fact that conductivity changes responsible for NDR can involve electronic, thermal or a combination of electronic and thermal processes, and is further complicated by the possibility of material-specific phase transitions and non-uniform current distributions due to electroforming or current constriction/bifurcation[12-18]. Despite these complexities, several studies have demonstrated that continuous S-type NDR in amorphous binary metal oxides can be well modelled by a temperature-dependent transport model, such as Poole-Frenkel conduction, and local Joule heating[19-23]. Indeed, Gibson recently generalized this understanding to show that S-type NDR can arise from any conduction mechanisms that depends superlinearly on temperature[24]. His study also provided a basis for correlating particular NDR characteristics with material properties, thereby providing a basis for engineering devices with specific characteristics.

While the above studies provide a foundation for understanding S-type NDR characteristics, they do not explain the diverse range of other NDR characteristics exhibited by thin film oxides, such as the discontinuous snap-back response observed in oxides such as $NbO_x$[25], $TaO_x$[26], $SiO_x$[27]. This is analogous to a similar response observed in chalcogenide glasses[28] and is characterised by abrupt, hysteretic voltage changes during bidirectional current sweeps. Devices exhibiting more complex combinations of continuous and snap-back characteristics have also been reported and have the potential to offer new functionality for neuromorphic computing[3]. In an attempt to understand the origin of such characteristics, Kumar et. al.[25] employed in-situ temperature mapping of amorphous $NbO_x$ devices that exhibited both S-type and snap-back characteristics. The onset of the continuous S-type NDR was found to occur at ~400 K, consistent with Poole-Frenkel conduction, while the snap-back characteristic was found to occur at ~1000 K and was attributed to the known insulator-metal-transition (IMT) in $NbO_2$[29-31]. However, the latter mechanism fails to account for similar snap-back behaviour in oxides such as $TaO_x$ and has been questioned by Goodwill et. al.[26] based on finite-element modelling of $TaO_x$ and $VO_x$ devices that reproduces the snap-back response without recourse to a material specific phase transition.

In this Article, we introduce a material independent model of current-controlled NDR that addresses the about controversy and explains a diverse range of discrete and compound NDR characteristics measured in $NbO_x$-based devices. This is achieved by explicitly accounting for a non-uniform current distribution in the device and its impact on the effective circuit of the device under current-controlled testing. Our results clearly define criteria for achieving continuous S-type and snap-back characteristics and further show that these serve as the fundamental building blocks for other compound NDR characteristics with higher complexity.

**Device structure and switching characteristics**

The device characteristics presented herein are derived from a range of studies by our group and are selected to demonstrate specific characteristics and functionality (see Supplementary Information Table S1). The devices are based on $NbO_x$/Pt cross point structures with Pt/Ti or Pt/Nb top electrodes (Fig. 1a) and were fabricated using standard photolithographic processes described elsewhere[32].

Switching characteristics were simulated using a lumped element model of an archetype threshold switch (memristor). For simplicity, the electrical conductivity was assumed to have a power law dependence on temperature such that the device resistance is given by:

$$R_m = R_0 (\frac{T_m}{T_{amb}})^{-\gamma}$$

where $T_m$ and $T_{amb}$ denote the temperature of the electrically active region and the ambient environment, $R_0$ is the resistance of the active region at T=$T_{amb}$ and $\gamma$ is an exponent that determines the temperature dependence. Gibson[24] has shown that a superlinear temperature dependence is required to produce current controlled NDR so that $\gamma > 1$. The dynamic behaviour of the memristor is then described by Newton's law of cooling:

$$\frac{dT_m}{dt} = \frac{I_m^2 R_m}{C_{th}} - \frac{\Delta T}{R_{th} C_{th}}$$

where $R_{th}$ and $C_{th}$ are the thermal resistance and the thermal capacitance of the device, and $\Delta T$ is the temperature difference between the $T_m$ and $T_{amb}$. This model was implemented as a circuit model using LT-Spice and details of the model and associated parameters are given in the Supplementary Information (see Section 2). Selected simulations of individual and compound NDR characteristics were also performed using the Poole-Frenkel conduction mechanism (see Section 2 of the Supplementary information).

Figs. 1b and 1c show representative switching characteristics that highlight the threshold switching and NDR responses observed under voltage and current-controlled testing, respectively. Such characteristics are the result of an increase in device conductance induced by local Joule heating and are accurately reproduced by lumped element models such as the one described above[22,28]. In contrast, Fig. 1d shows the current-voltage characteristics of a device that exhibits discontinuous I-V characteristics under current-controlled testing, referred to as a "snap-back" response. This behaviour is a generic response observed in many oxides but is not predicted by the archetype lumped element model and clearly involves physical processes not included in the model.

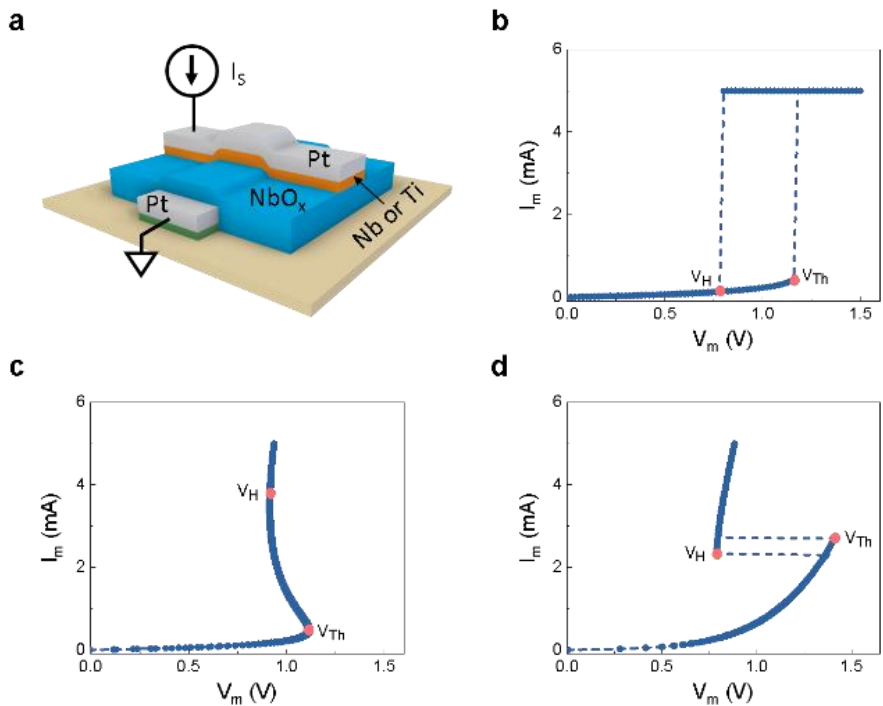

**Fig. 1 | Device structure and current-voltage characteristics**. **a,** Schematic of metal-oxide-metal cross-point device structures used in this study. **b,** Measured current-voltage characteristic under voltage-controlled operation. **c,d,** Measured current-voltage characteristics of devices under current-controlled operation showing continuous S-type and discontinuous ("snap-back") behaviour, respectively.

**Underlying physical mechanism**

Here we introduce a new model that addresses the above shortcoming and provides a comprehensive understanding of discrete and compound NDR characteristics. This is achieved by explicitly accounting for a non-uniform current (temperature) distribution in the oxide film and its impact on the effective circuit of the device. The response of the oxide film is then heterogeneous, with the hottest region being the first to undergo current-controlled NDR and the cooler regions acting as a parallel resistance. For a strongly peaked temperature distribution, as is the case for filamentary conduction, the distribution can be represented by a core-shell structure comprising the filamentary conduction path (core) and a parallel resistance due to the surrounding film (shell) [26], as shown in Fig. 2 (top). The total device current is then distributed between these two regions, with the magnitude of the parallel resistance depending on the conductivity of the film and the temperature distribution in the device. Although the total device current remains constrained under current controlled operation, the current ratio in the core and shell regions is free to respond to local conductivity changes and can facilitate a sudden reduction in core resistance if the resistance of the shell is sufficiently low. As discussed below, this simple two-zone, core-shell model is sufficient to explain the diverse range of switching phenomena observed in oxide devices, including the snap-back response shown in Fig. 1d.

As the core and shell regions represent different areas of the same oxide film, their electrical conductivity is expected to be governed by the same physical mechanism. As a consequence the structure can be represented by two parallel archetype memristors (as defined above), one representing the high-conductivity filament, with resistance $R_{m1}(I_c)$, and the other the parallel conduction through the surrounding film, with resistance $R_{m2}(I_s)$, as shown in Fig. 2 (top). The total device resistance is then $R_{dev} = \frac{R_{m1} R_{m2}}{R_{m1} + R_{m2}}$ where both $R_{m1}(I_c)$ and $R_{m2}(I_s)$ can be negative within specific current ranges.

For a strongly localised current distribution (e.g. filamentary conduction) the core region can be represented by an archetype memristor and the surrounding shell by a fixed (i.e. temperature independent) resistor of resistance $R_S = R_{m2}(I = 0)$. In this case the switching characteristics are determined by the relative magnitudes of the shell resistance $R_S$ and the maximum negative differential resistance of the core region $R_{NDR} = max \left|\frac{dV}{dI}\right|_{NDR}$. When the parallel resistance $R_S$ is much larger than $R_{NDR}$ the source current, $I_s$, flows predominantly through the core and gives rise to a continuous S-type NDR characteristics similar to that of the archetype memristor (Fig. 2a). In contrast, when $R_S$ is less than $R_{NDR}$ the current flows mainly through the shell region and can have a lower current density than the core due to their relative areas. In this case the shell resistance effectively acts a current divider that enables the core to respond to the positive feedback created by local Joule heating while maintaining constant total device current. The resulting increase in core conductivity is then analogous to the threshold switching response shown in Fig. 1b and produces the sudden reduction in voltage that characterizes the snap-back response in Fig. 2b. In essence, the current source and parallel shell resistance act as a voltage source for the core memristor and it undergoes voltage-controlled threshold switching when the device voltage exceeds it threshold value, $V_{Th}$. More generally, it is necessary to consider that both the core and shell regions have temperature dependent conductivities that can be represented by archetype memristors, as shown in Fig. 2c. In this case the resulting NDR characteristics reflect the relative magnitudes and temperature dependencies of the two regions and include a wide range of compound characteristics, an example of which is illustrated in Fig. 2c.

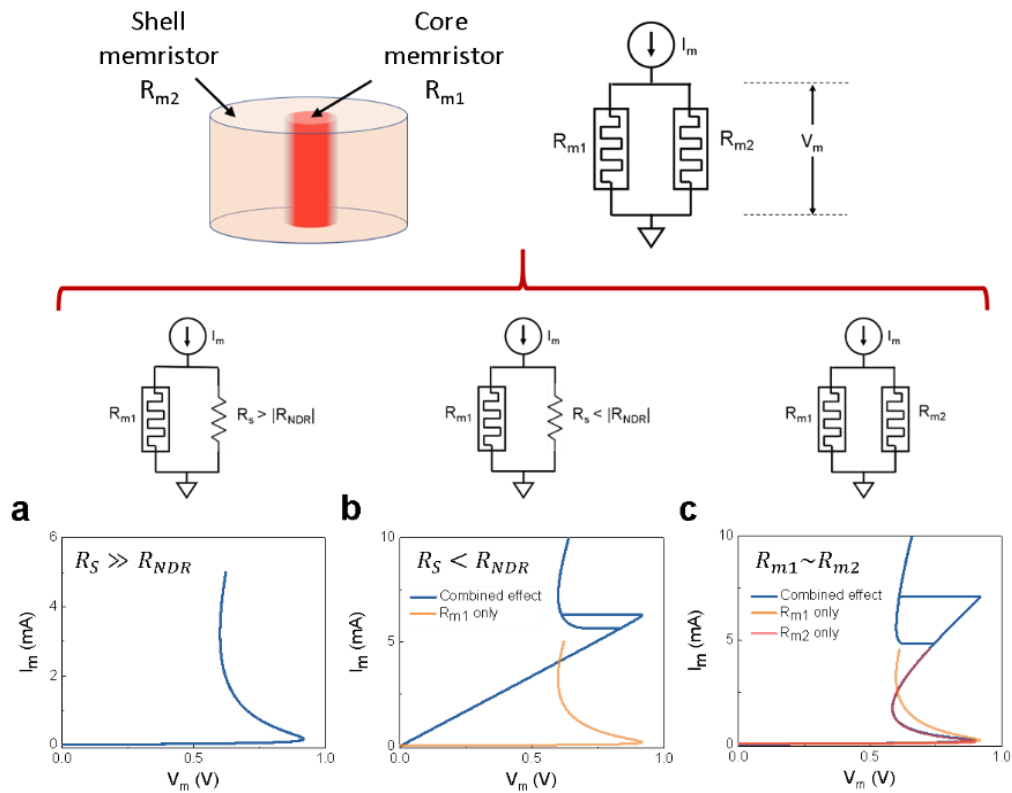

**Fig. 2 | Lumped element models of devices with different core-shell resistance ratios and the calculated switching response. a,** The shell resistance is constant and greater than the negative differential resistance of the core. **b,** The shell resistance is constant and lower than the negative differential resistance of the core. The orange coloured solid line shows the current-voltage curve of $R_{m1}$ without the parallel $R_s$. **c,** The shell resistance changes with current and is treated as a second memristor. The red and orange coloured solid lines represent the individual current-voltage characteristics of $R_{m1}$ and $R_{m2}$.

Fig. 3 shows more quantitative analysis of these dependencies, including calculated device temperatures (see Section 2 of the Supplementary Information for details of the model). Here the core region of the devices is represented by the same archetype memristor while the shell region is variously represented by a fixed resistor or a second memristor. Fig. 3a shows the differential resistance, dV/dI, derived from the current-voltage characteristics of the core memristor, showing that the absolute maximum of the NDR is ~ 380 Ω at a current of ~315 µA. Based on the above discussion this resistance sets the criterion for the S-type and snap-back responses, with the S-type NDR observed for shell-resistances greater than this value and snap-back NDR observed for shell-resistances less than this value. In this context, the current-voltage characteristic of the isolated core memristor corresponds to the former case with $R_S \gg R_{NDR}$ (i.e. $R_S = \infty$) and exhibits the expected continuous S-type NDR characteristics, as shown in Fig. 3b. The onset temperature in this case is predicted to be ~450 K, consistent with simulations based on Poole-Frenkel conduction[22] and experimental observations[25]. The counter example is shown if Fig. 3c which depicts the characteristics of the same device but with a parallel shell resistance of 150 Ω, i.e. $R_S < R_{NDR}$. In this case the device exhibits a hysteretic snap-back response similar to that observed experimentally (e.g. Fig. 1d). Significantly, the onset temperature for the snap-back response is similar to that for the S-type response, as expected from their common origin. Indeed, this clearly demonstrates that the snap-back response results from current redistribution between the core and shell regions of the device rather than a material-specific phase transition. Finally, Fig. 3d shows corresponding calculations for a device in which the temperature dependencies of both the core and shell regions are considered. In this case, the shell region is also represented by an archetype memristor albeit with slightly different resistance and temperature dependence than the core. The resulting compound characteristics reflect the relative resistances and temperature dependencies of the core and shell regions and represent only one of many compound responses, as discussed below.

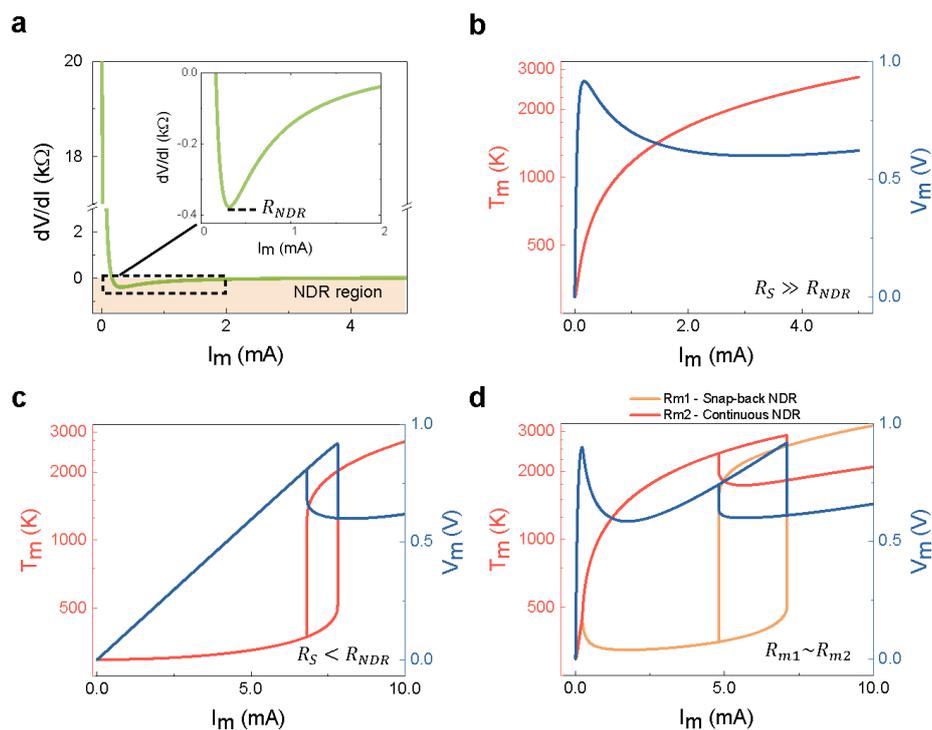

**Fig. 3 | Dynamic resistance and simulated temperature profile. a,** The dynamic resistance, $\left|\frac{dV}{dI}\right|$, plotted as a function of $I_m$. The inset shows an expanded area of the region where $R_{NDR}$ is the maximum value of negative differential resistance $\left|\frac{dV}{dI}\right|$. **b,** Simulated device temperature of continuous S-type NDR with corresponding current-voltage curve with $R_S \gg R_{NDR}$. The device temperature profile shows a gradual and continuous increase with increasing current. **c,** Plot of device temperature and the quasi-static current-voltage curve as a function of current for snap-back NDR with $R_S < R_{NDR}$. The onset temperature remains relatively low at ~450 K and is followed by an abrupt increase in temperature. **d,** Simulated temperature profile of two active regions for the compound NDR behaviour which consists of one continuous S-type and one snap-back NDRs.

**Engineering devices with compound characteristics**

Despite the apparent simplicity of the lumped element core/shell model it has proven sufficient to reproduce a broad range of experimentally observed compound NDR characteristics in our $NbO_x$-based devices. This is highlighted in Fig. 4 which compares a range of predicted and experimentally observed characteristics. For example, Fig. 4a illustrates a case where current flow is initially concentrated in the core region to produce a continuous S-type NDR response at low currents. As the current continues to increase, the temperature of the shell also increases to the point where it exhibits S-type NDR but in this case the core resistance is too high to meet the requirements for a snap-back response. The device therefore exhibits two continuous S-type NDRs. In contrast, Fig. 4b shows a case where the core resistance is sufficient low at high currents to meet the requirements for a snap-back response in the shell region. This produces one continuous S-type NDR and a snap-back NDR, similar to the reported behaviour[25]. The final example, Fig. 4c, corresponds to a case where the conditions for a snap-back response are met by both the core and shell regions, thereby resulting in a double snap-back response. Significantly, each of these characteristics has been observed experimentally in discrete devices, as shown in Figs. 4d-f. Experimentally, the characteristics are determined by the film stoichiometry, device structure and operating conditions (i.e. current compliance during electroforming or maximum operating current).

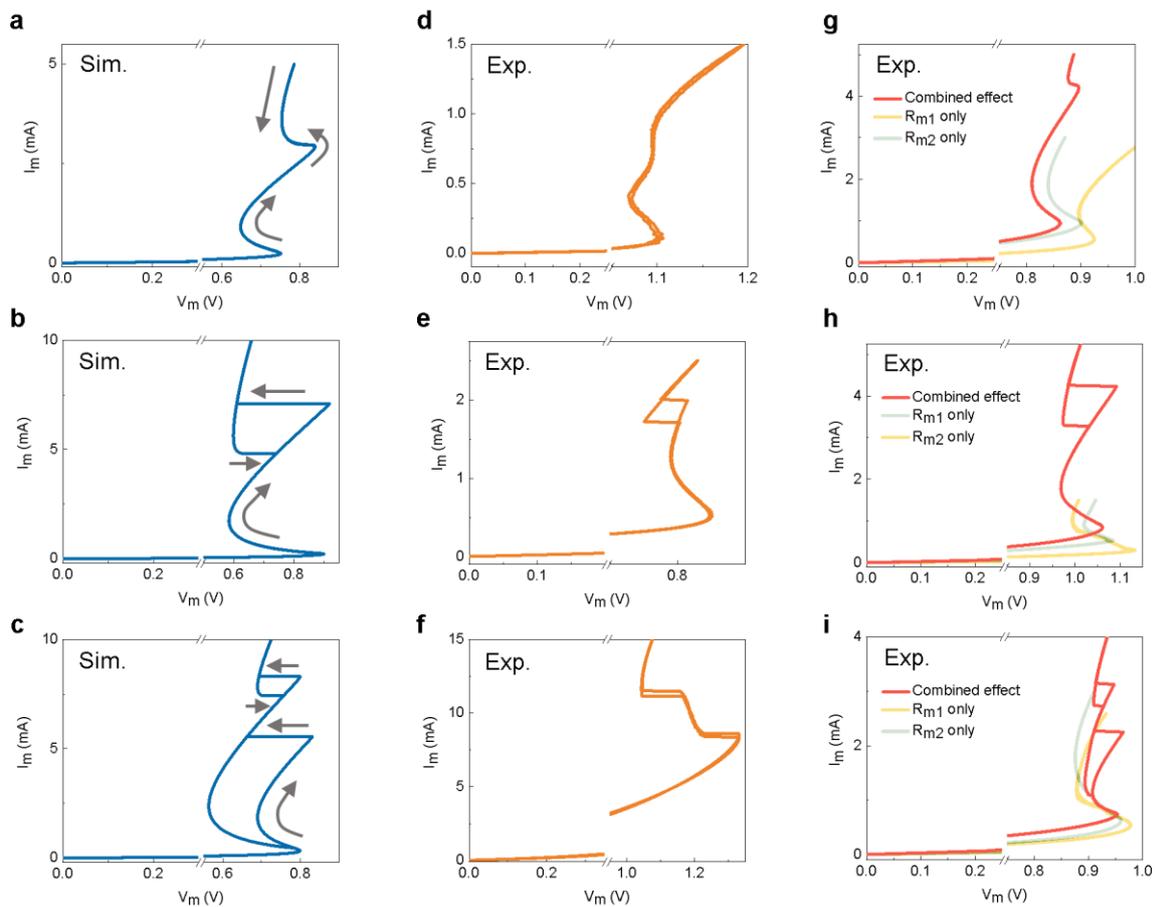

**Fig. 4 | Compound NDR characteristics in individual and coupled devices.** Simulation predicted compound NDR characteristics for two parallel memristors that exhibit continuous S-type NDR, including: **a**, the co-existence of two continuous NDR regions. **b**, one continuous S-type NDR with one snap-back NDR **c**, two snap-back NDR regions. **d-f**, Experimentally measured compound NDR characteristics for individual $NbO_x$ devices. **g-i,** Experimentally measured compound NDR characteristics for coupled $NbO_x$ devices formed by combining two discrete devices in parallel; The response of the individual devices is included for reference. Note: all the compound behaviours have also been simulated and predicted with the Poole-Frenkel conduction (Fig. S4 in the Supplementary Information).

It is interesting to note that the above discussion does not explicitly require that the shell resistance $R_{m2}(I_s)$ be a component of the active device. It could instead be a separate circuit element such as an external resistor or a second memristor. Indeed, when an external parallel resistor is added to a device exhibiting continuous S-type NDR the characteristics transition from continuous to snap-back NDR behaviour as the resistor is varied to satisfy the criteria discussed above (see Section 1 of the Supplementary Information). More generally, the full complexity of the predicted characteristics (Figs. 4a-c) was reproduced by combining two discrete S-type devices in parallel, as shown in Figs. 4g-i. This is particularly significant because the realization and reproducibility of specific composite NDR characteristics in individual devices can be challenging due to limited control of the core-shell structure, particularly for nanoscale devices where the formation of a distinct core-shell structure is less likely. The ability to control the switching characteristics by combining discrete S-type NDR devices addresses these issues and provides additional flexibility for designing devices with specific functionality.

**Designing composite behaviour with higher complexity**

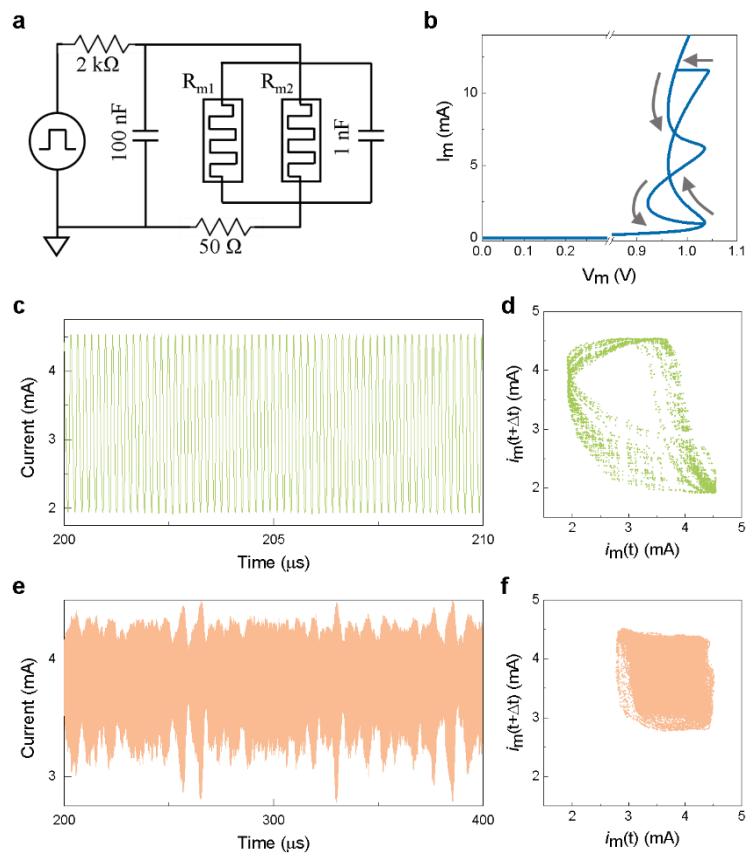

**Fig. 5 | Simulations of complex NDR characteristics and associated dynamical behaviour. a,** Schematic of the oscillator circuit configuration used in the simulation. **b,** Two memristors that exhibit continuous S-type NDR, an unconventional complex compound NDR behaviour is predicted in simulation which includes a continuous S-type NDR followed by a snap-back response in the forward current sweep and two continuous S-type NDRs regions in the reverse sweep. **c-d,** Current oscillations in time domain and the corresponding recurrence plot (Δt = 0.6 μs) at an applied DC voltage pulse of 7.5 V. **e-f**, Current oscillations in time domain and the corresponding recurrence plot (Δt = 0.6 μs) with a DC voltage pulse input of 8.7V.

From the above understanding it is also possible to understand and engineer devices with unique I-V characteristics and functionality. As an example, Fig. 5 shows the equivalent circuit and predicted characteristics of a novel compound device with higher complexity. The I-V characteristics show combined S-type and snap-back responses during a forward current-sweep and two continuous S-type responses during a reverse current sweep. The dynamic behaviour of this compound NDR response was studied using a Pearson-Anson oscillator circuit (Fig. 5a) with a load resistor of 2 k$\Omega$, a 50 $\Omega$ monitoring resistor, an external parallel 100 nF capacitor and an intrinsic capacitance of 1 nF (see Section 3 of the Supplementary Information for details of the model). The results show two types of oscillation dynamics. For an input voltage pulse of 7.5 V the device exhibits periodic oscillations with high peak-to-peak amplitude, as shown in Fig. 5c and validated in the recurrence plot of Fig. 5d. In contrast, when the input voltage is increased to ~8.7 V the system exhibits irregular or chaotic-like characteristics with lower peak-to-peak amplitude (Fig. 5e). The irregularity of the oscillation is confirmed by the recurrent plot (Fig. 5f). Complex and compound NDR behaviours with even higher order of complexity were also predicted with triple memristors in parallel (see Section 4 of the Supplementary Information). We note that characteristics similar to those depicted in Fig. 4a have been realized experimentally and are the subject of further study. However, this example serves to illustrate the potential of the simple core-shell model for exploring devices with novel functionality.

**Conclusions**

A material-independent model of current-controlled negative differential resistance was introduced and shown to explain a broad range of reported switching characteristics, including the so-called "snap-back" response observed in transition metal-oxides. The model assumed a non-uniform current distribution in the film and took account of its effect on the equivalent circuit of the device. For highly localized current distributions this amounted to approximating the film by a core-shell structure in which the core was treated as an archetype memristor and the shell was approximated as a parallel resistor. In this case the continuous and snap-back switching modes were shown to depend on the relative resistances of the core and shell regions. The model was then extended to include the case where both the core and shell have memristive behaviour. This demonstrated that the continuous S-type and snap-back responses serve as fundamental building blocks for NDR characteristics with higher complexity and novel functionality for future electronics and emerging computing paradigm.

**References**

1. Yu, S. Neuro-inspired computing with emerging nonvolatile memorys. *Proc. IEEE* **106**, 260-285 (2018).
2. Liu, X., Li, S., Nandi, S. K., Venkatachalam, D. K. & Elliman, R. G. Threshold switching and electrical self-oscillation in niobium oxide films. *J. Appl. Phys.* **120**, 124102 (2016).
3. Kumar, S., Strachan, J. P. & Williams, R. S. Chaotic dynamics in nanoscale $NbO_2$ Mott memristors for analogue computing. *Nature* **548**, 318 (2017).
4. Sharma, A. A., Li, Y., Skowronski, M., Bain, J. A. & Weldon, J. A. High-Frequency $TaO_x$-Based Compact Oscillators. *IEEE Trans. Elec. Dev.* **62**, 3857-3862 (2015).
5. Mian, M. S., Okimura, K. & Sakai, J. Self-oscillation up to 9 MHz based on voltage triggered switching in $VO_2$/TiN point contact junctions. *J. Appl. Phys.* **117**, 215305 (2015).
6. Parihar, A., Shukla, N., Jerry, M., Datta, S. & Raychowdhury, A. Vertex coloring of graphs via phase dynamics of coupled oscillatory networks. *Sci. Rep.* **7**, 911 (2017).
7. Zhao, B. & Ravichandran. Low-Power Microwave Relaxation Oscillators Based on Phase-Change Oxides for Neuromorphic Computing. *Phys. Rev. Appl* **11**, 014020 (2019).
8. Lappalainen, J., Mizsei, J. & Huotari, M. Neuromorphic thermal-electric circuits based on phase-change $VO_2$ thin-film memristor elements. *J. Appl. Phys.* **125**, 044501 (2019).
9. Pickett, M. D. & Williams, R. S. Phase transitions enable computational universality in neuristor-based cellular automata. *Nanotechnology* **24**, 384002 (2013).
10. Yi, W. *et al.* Biological plausibility and stochasticity in scalable $VO_2$ active memristor neurons. *Nat. Commun.* **9**, 4661 (2018).
11. Pickett, M. D., Medeiros-Ribeiro, G. & Williams, R. S. A scalable neuristor built with Mott memristors. *Nat. Mater.* **12**, 114 (2013).
12. Kumar, S. & Williams, R. S. Separation of current density and electric field domains caused by nonlinear electronic instabilities. *Nat. Commun.* (2018).
13. Ridley, B. Specific negative resistance in solids. *Proc. Phys. Soc. Lond.* **82**, 954 (1963).
14. Wang, M. *et al.* S-Type Negative Differential Resistance in Semiconducting Transition-Metal Dichalcogenides. *Adv. Elec. Mater.*, 1800853 (2019).
15. Ovshinsky, S. R. Reversible electrical switching phenomena in disordered structures. *Phys. Rev. Lett.* **21**, 1450 (1968).
16. Czubatyj, W. & Hudgens, S. J. Thin-film Ovonic threshold switch: Its operation and application in modern integrated circuits. *Electron. Mater. Lett.* **8**, 157-167 (2012).
17. Park, J. H. *et al.* Measurement of a solid-state triple point at the metal–insulator transition in VO2. *Nature* **500**, 431 (2013).
18. Imada, M., Fujimori, A. & Tokura, Y. Metal-insulator transitions. *Rev. Mod. Phys.* **70**, 1039 (1998).
19. Li, S., Liu, X., Nandi, S. K. & Elliman, R. G. Anatomy of filamentary threshold switching in amorphous niobium oxide. *Nanotechnology* **29**, 375705 (2018).
20. Funck, C. *et al.* Multidimensional simulation of threshold switching in $NbO_2$ based on an electric field triggered thermal runaway model. *Adv. Elec. Mater.* **2**, 1600169 (2016).
21. Goodwill, J. M., Sharma, A. A., Li, D., Bain, J. A. & Skowronski, M. Electro-thermal model of threshold switching in $TaO_x$-based devices. *ACS Appl. Mater. Interf.* **9**, 11704-11710 (2017).

| | |
|---|---|
| 22 | Slesazeck, S. *et al.* Physical model of threshold switching in NbO$_2$ based memristors. *RSC Adv.* **5**, 102318-102322 (2015). |
| 23 | Gibson, G. A. *et al.* An accurate locally active memristor model for S-type negative differential resistance in NbO$_x$. *Appl. Phys. Lett.* **108**, 023505 (2016). |
| 24 | Gibson, G. A. Designing Negative Differential Resistance Devices Based on Self-Heating. *Adv. Funct. Mater.* **28**, 1704175 (2018). |
| 25 | Kumar, S. *et al.* Physical origins of current and temperature controlled negative differential resistances in NbO$_2$. *Nat. Commun.* **8**, 658 (2017). |
| 26 | Goodwill, J. M. *et al.* Spontaneous current constriction in threshold switching devices. *Nat. Commun.* **10**, 1628 (2019). |
| 27 | Chen, T., Tse, M., Sun, C., Fung, S. & Lo, K. Snapback behaviour and its similarity to the switching behaviour in ultra-thin silicon dioxide films after hard breakdown. *J. Phys. D. Appl. Phys.* **34**, L95 (2001). |
| 28 | Suntola, T. On the mechanism of switching effects in chalcogenide thin films. *Solid State Electron.* **14**, 933-938 (1971). |
| 29 | Janninck, R. & Whitmore, D. Electrical conductivity and thermoelectric power of niobium dioxide. *J. Phys. Chem. Solids* **27**, 1183-1187 (1966). |
| 30 | Adler, D. Mechanisms for metal-nonmental transitions in transition-metal oxides and sulfides. *Rev. Mod. Phys.* **40**, 714 (1968). |
| 31 | Eyert, V. The metal-insulator transition of NbO$_2$: An embedded Peierls instability. *EPL* **58**, 851 (2002). |
| 32 | Nath, S. K., Nandi, S. K., Li, S. & Elliman, R. G. Detection and spatial mapping of conductive filaments in metal/oxide/metal cross-point devices using a thin photoresist layer. *Appl. Phys. Lett.* **114**, 062901 (2019). |

## Acknowledgements

This work was partly funded by the Australian Research Council (ARC) and Varian Semiconductor Equipment/ Applied Materials through an ARC Linkage Project Grant: LP150100693. We would like to acknowledge access to NCRIS facilities at the ACT node of the Australian Nanotechnology Fabrication Facility (ANFF) and the Australian Facility for Advanced ion-implantation Research (AFAiiR), and thank Dr Tom Ratcliff for comments and feedback on the manuscript.

## Competing interests

The authors declare no competing interests.

## Author contributions

S.L., X. L. and R.G. E. conceived the concept. S.L. and X.L. performed the simulations of the models. S.L., S.K.N and S.K.N. fabricated the devices and performed experimental measurements. S.L. and R.G.E. wrote the manuscript. All authors discussed the results and implications and commented on the manuscript at all stages.

# Supplementary Information

## 1. Device Controllability of NDR

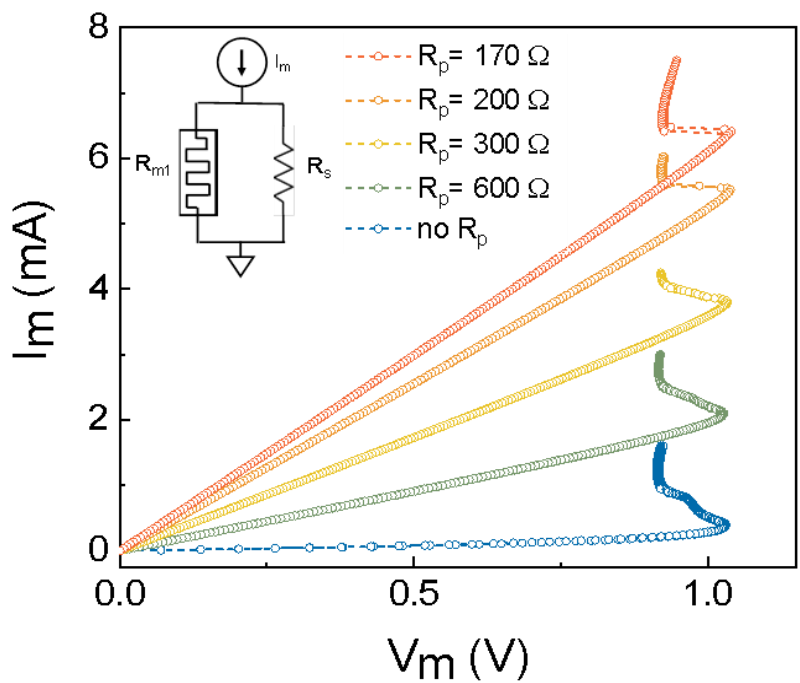

**Figure S1.** Current-voltage plots of NDR tunability via an external resistor in parallel with a crossbar NbO$_x$ memristor. The external resistance was applied in the range from 600 Ω to 170 Ω with transitions from continuous S-type NDR to snap-back NDR.

**Table S1.** Measured device structures.

| Data | Top electrode thickness (nm) | NbO$_x$ (nm) | NbO$_x$ composition, x | Bottom electrode and thickness (nm) | Device dimension |
|---|---|---|---|---|---|
| Fig. 1e, Fig. 4e | Pt(25)/Nb(10) | 50 | 2.15±0.05 | Pt (25) | 10 μm x 10 μm |
| Fig. 1d | Pt(25)/Nb(10) | 50 | 2.60±0.05 | Pt (25) | 10 μm x 10 μm |
| Fig 4d, 4g-i | Pt(25)/Ti(5) | 50 | 2.60±0.05 | Pt (25) | 10 μm x 10 μm |
| Fig 4f | Pt(50) | 50 | 2.60±0.05 | TiN(50) | 50 μm diameter CV pad |

## 2. SPICE Models for archetype threshold switching memristor

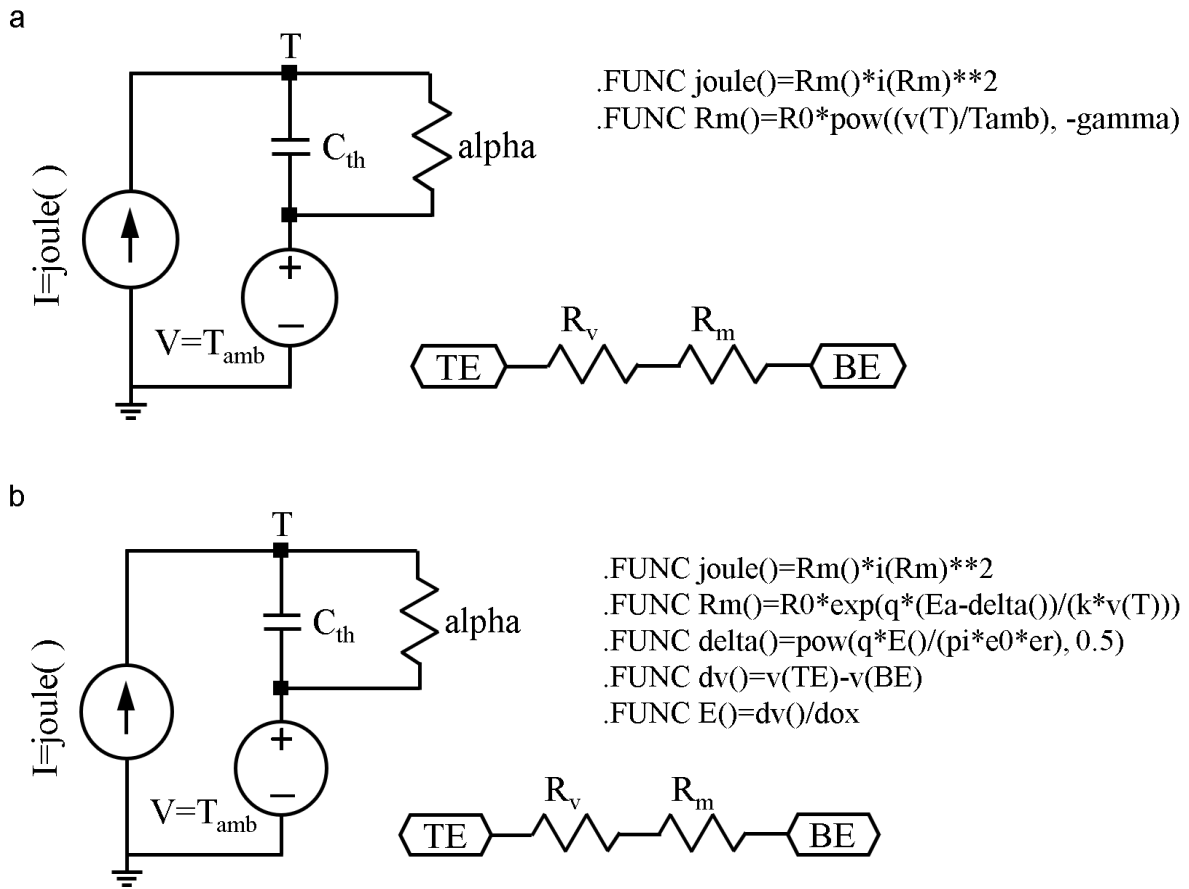

**Figure S2. a**, SPICE model for threshold switching memristor based on Joule heating and an electrical conductivity with a power-law dependence (equation 1 and 2). **b,** SPICE model for threshold switching memristor based on Joule heating and standard Poole-Frenkel conduction.

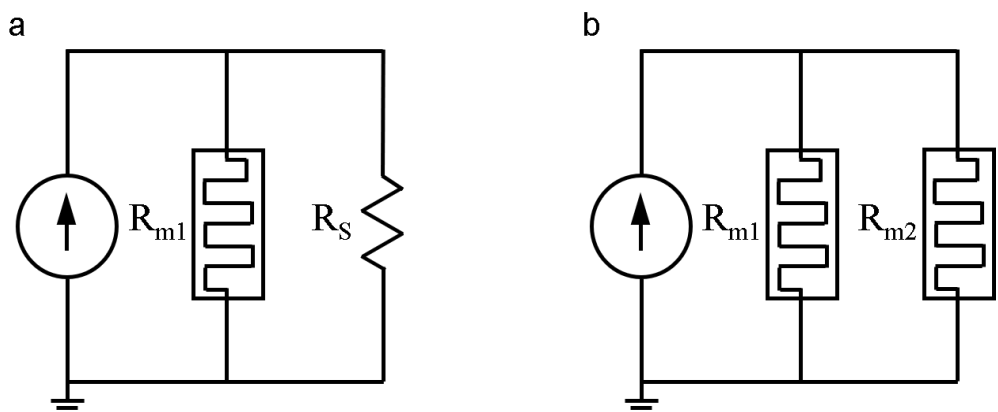

**Figure S3. a**, SPICE equivalent circuit of a core-shell structure for one active memristor and a passive parallel resistor in parallel. **b**, SPICE equivalent circuit of a core-shell structure for two active memristors in parallel.

**Table S2.** Memristor parameters used in simulation for the generic conductivity model.

| Model Parameters (Unit) | Symbol | $R_{m1}$ Value | $R_{m2}$ Value |
|---|---|---|---|
| Thermal capacitance (J·K$^{-1}$) | $C_{th}$ | $1\times10^{-15}$ | $1\times10^{-15}$ |
| Resistance prefactor (kΩ) | $R_0$ | 15 to 30 | 15 to 30 |
| Thermal resistance (K·W$^{-1}$) | α | $1.6\times10^6$ to $1.8\times10^6$ | $1.1\times10^6$ to $1.6\times10^6$ |
| Ambient temperature (K) | $T_{amb}$ | 296.15 | 296.15 |
| Superlinear coefficient | γ | 2 to 3 | 2 to 3.5 |
| Shell resistance (Ω) | $R_s$ | 150 | N/A |
| Metallic state resistance (Ω) | $R_v$ | 50 to 200 | 70 to 100 |

**Table S3.** Memristor parameters used in simulation for the Poole-Frenkel conduction model.

| Model Parameters (Unit) | Symbol | $R_{m1}$ Value | $R_{m2}$ Value |
|---|---|---|---|
| Thermal capacitance (J·K$^{-1}$) | $C_{th}$ | $1\times10^{-15}$ | $1\times10^{-15}$ |
| Resistance prefactor (Ω) | $R_0$ | 25 to 75 | 25 |
| Thermal resistance (K·W$^{-1}$) | α | $1.3\times10^5$ to $1\times10^6$ | $2.2\times10^5$ to $1\times10^6$ |
| Ambient temperature (K) | $T_{amb}$ | 296.15 | 296.15 |
| Shell resistance (Ω) | $R_s$ | 100 | N/A |
| Metallic state resistance (Ω) | $R_v$ | 0 | 0 |
| Activation Energy (eV) | $E_a$ | 0.215 | 0.215 to 0.225 |
| Boltzmann constant (J·K$^{-1}$) | $k_B$ | $1.38\times10^{-23}$ | $1.38\times10^{-23}$ |
| Elementary charge (C) | e | $1.6\times10^{-19}$ | $1.6\times10^{-19}$ |
| Vacuum permittivity (F·m$^{-1}$) | $\varepsilon_0$ | $8.85\times10^{-12}$ | $8.85\times10^{-12}$ |
| Relative permittivity of the threshold switching volume | $\varepsilon_r$ | 45 | 45 |
| Thickness of the threshold switching volume (nm) | $t_v$ | 20 to 30 | 50 to 125 |

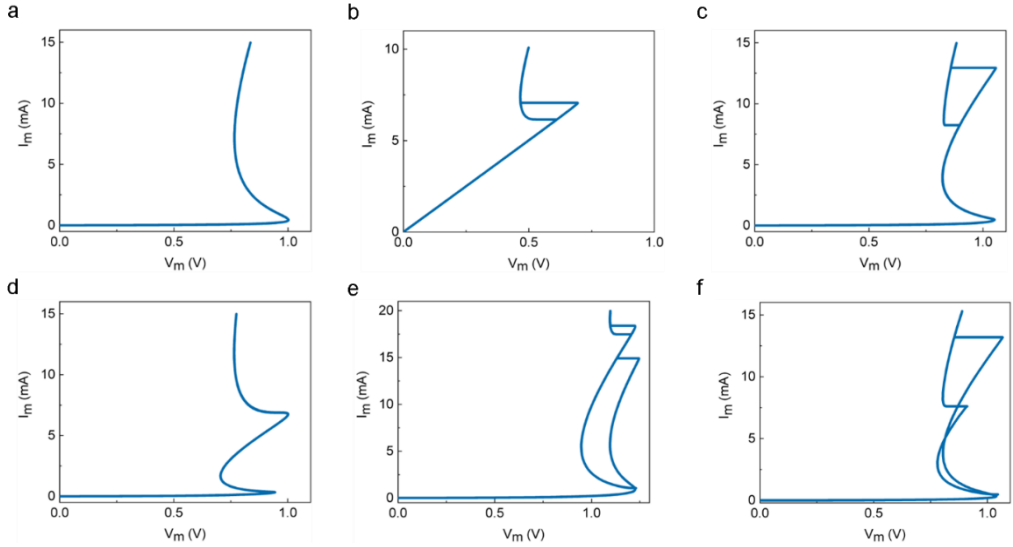

**Figure S4. a,** simulated S-type NDR response. **b,** Simulated discontinuous snap-back characteristics. the co-existence of two continuous NDR regions. **c,** Co-existence of one continuous S-type NDR with one snap-back NDR **d,** Predicted compound characteristics with two snap-back NDR regions. **e,** Composite behaviour with two snap-back NDR regions. **f,** Composite behaviour with higher complexity.

### 3. SPICE Model for dynamical behaviour

**Table S4.** Memristor parameters used in dynamical simulation based on the Poole-Frenkel conduction model.

| Model Parameters (Unit) | Symbol | $R_{m1}$ Value | $R_{m2}$ Value |
|---|---|---|---|
| Thermal capacitance (J·K$^{-1}$) | $C_{th}$ | $2.5 \times 10^{-14}$ | $2.5 \times 10^{-14}$ |
| Resistance prefactor (Ω) | $R_0$ | 47 | 110 |
| Thermal resistance (K·W$^{-1}$) | α | $1.5 \times 10^5$ | $4.2 \times 10^5$ |
| Ambient temperature (K) | $T_{amb}$ | 296.15 | 296.15 |
| Metallic state resistance (Ω) | $R_v$ | 0 | 0 |
| Activation Energy (eV) | $E_a$ | 0.204 | 0.202 |
| Boltzmann constant (J·K$^{-1}$) | $k_B$ | $1.38 \times 10^{-23}$ | $1.38 \times 10^{-23}$ |
| Elementary charge (C) | e | $1.6 \times 10^{-19}$ | $1.6 \times 10^{-19}$ |
| Vacuum permittivity (F·m$^{-1}$) | $\varepsilon_0$ | $8.85 \times 10^{-12}$ | $8.85 \times 10^{-12}$ |
| Relative permittivity of the threshold switching volume | $\varepsilon_r$ | 45 | 45 |
| Thickness of the threshold switching volume (nm) | $t_v$ | 19 | 24 |

## 4. Predicted compound NDR behaviour with triple memristors in parallel

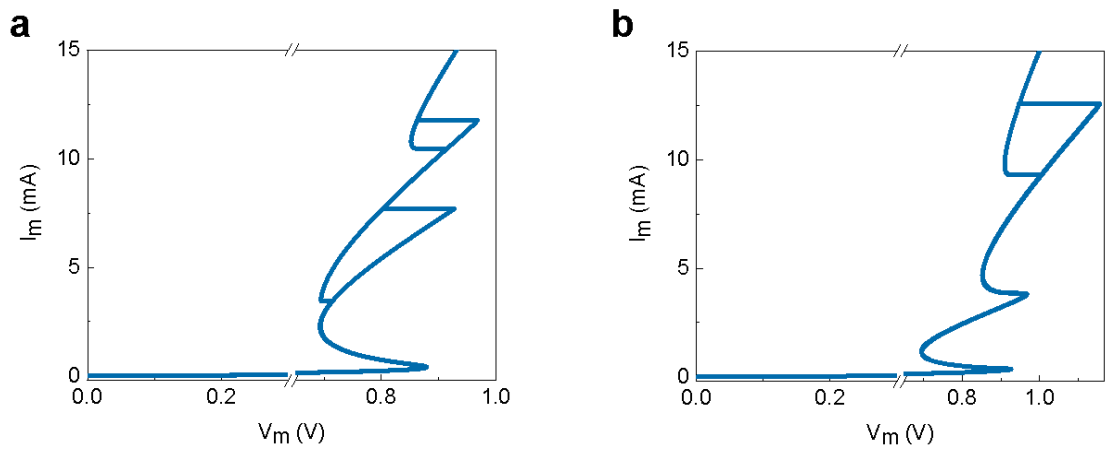

**Figure S5. a-b**, Examples of simulated compound NDR characteristics with three memristors in parallel.